\documentstyle[12pt]{article}

\begin{document}

{\it University of Shizuoka}

\hspace*{10.5cm} {\bf US-00-03}\\[-.3in]

\hspace*{10.5cm} {\bf May 2000}\\[.3in]

\vspace*{.4in}

\begin{center}

{\Large\bf  Quark and Lepton Mass Matrices  \\[.1in]
with a Cyclic Permutation Invariant Form} \\[.3in]

{\bf Yoshio Koide}\footnote{
E-mail: koide@u-shizuoka-ken.ac.jp} \\

Department of Physics, University of Shizuoka \\ 
52-1 Yada, Shizuoka 422-8526, Japan \\[.1in]

\vspace{.3in}

{\large\bf Abstract}\\[.1in]

\end{center}

\begin{quotation}
As an attempt to give an unified description of quark and lepton mass matrices 
$M_f$, the following mass matrix form is proposed: the form of the mass 
matrices are invariant under a cyclic permutation 
$(f_1{\rightarrow}f_2, f_2{\rightarrow}f_3,f_3{\rightarrow}f_1)$ 
among the fermions $f_i$.  
The model naturally leads to 
the maximal mixing between $\nu_{\mu}$ and $\nu_{\tau}$, and with an additional
 ansatz, it leads to the well-satisfied relations $|V_{us}|\simeq{\sqrt{m_d/m_s}}$ and $|V_{cb}|\simeq{\sqrt{m_d/m_b}}$.  
\end{quotation}

\vfill
%%%%%%%%%%%%%%%%%%%%%%%%%%%%%%%%%%%%%%%%%%%
\newpage
\section{Introduction}

The times of the phenomenological study of the quark mass matrices
have been over.
However, as far as the study of the neutrino mass matrix is concerned,
the phenomenological study  still have the meaning.
It is still important to investigate the unified description of
the quark and lepton mass matrices from the phenomenological point of
view.

In such a phenomenological study, the key to the unified description
is to find a fermion basis on which the quark and lepton mass matrices
at the unification energy scale take a simpler and beautiful form.
Concerning this point, recently, Takasugi and his collaborators 
\cite{Takasugi}  have discussed a neutrino mass matrix on a very 
interesting basis, which is described by a discrete symmetry Z$_3$.  
However, in their neutrino mass matrix model, 
the charged lepton mass matrix is given by a diagonal form, 
so that the model cannot give any predictions for charged lepton mass 
spectrum.  
And, at present, they have not applied their idea to the quark mass
matrices.
However, their basic idea seems to be highly promising for a unified 
description of quark and lepton mass matrices.

Stimulated by their idea, in the present paper, we investigate fermion 
mass matrices  with the form
$$
M=aE+bS(\theta), \eqno(1.1)
$$
$$
E={\frac{1}{\sqrt{3}}}
\left(\begin{array}{ccc}
1 & 0 & 0 \\
0 & 1 & 0 \\
0 & 0 & 1 
\end{array}\right),\ \ 
S(\theta)={\frac{1}{\sqrt{6}}}
\left(\begin{array}{ccc}
0 & e^{i\theta} & e^{-i\theta} \\
e^{-i\theta} & 0 & e^{i\theta} \\
e^{i\theta} & e^{-i\theta} & 0 \\
\end{array}\right),
\eqno(1.2)
$$
where the matrices $E$ and $S$($\theta$) have been normalized as 
Tr$E^2$=Tr$S^2$($\theta$)=1.  Although Takasugi and his collaborators 
\cite{Takasugi} 
have related the form (1.1) to a Z$_3$ symmetry, in this paper, 
we require that 
the mass matrix is invariant under a cyclic permutation 
($f_1{\rightarrow}f_2,f_2{\rightarrow}f_3,f_3{\rightarrow}f_1$) 
where $f_i$ are quarks and lepton fields ($f=u,d,\nu,e$).  The form 
(1.1) is the most general form which is invariant under the cyclic 
permutation and which is Hermitian.  The matrix $M$ given in Eq.~(1.1) 
is diagonalized by the tri-maximal mixing matrix \cite{tri-max}
$$
V_T\equiv{\frac{1}{\sqrt{3}}}
\left(\begin{array}{ccc}
1 & 1 & 1 \\
\omega & \omega^2 & 1 \\
\omega^2 & \omega & 1 
\end{array}\right),
\eqno(1.3)
$$
where $\omega=e^{2\pi i/3}$, as follows: 
$$
V_TMV_T^{\dagger}=
\left(\begin{array}{ccc}
m_1 & 0 & 0 \\
0 & m_2 & 0 \\
0 & 0 & m_3
\end{array}\right),
\eqno(1.4)
$$

$$
m_1={\frac{1}{\sqrt{3}}}a+{\frac{2}{\sqrt{6}}}b\cos\theta,
$$
$$
m_2={\frac{1}{\sqrt{3}}}a-{\frac{1}{\sqrt{6}}}b\cos\theta+
{\frac{1}{\sqrt{2}}}b\sin\theta,\eqno(1.5)
$$
$$
m_3={\frac{1}{\sqrt{3}}}a-{\frac{1}{\sqrt{6}}}b\cos\theta-
{\frac{1}{\sqrt{2}}}b\sin\theta,
$$
Note that any mass spectrum with three families can be described 
by the form given in Eq.~(1.1), because the three terms ($a$-, 
$b\cos\theta$-, and $b\sin\theta$-terms) in Eq.~(1.1) are 
transformed into the three diagonal matrices, diag$(1,1,1)$, 
diag$(2,-1,-1)$, and diag$(0,1,-1)$, respectively, which are 
independent of each other.  Therefore, we have 
the same number of parameters as the number of the observable 
quantities (mass values).  Besides, 
if the mass matrices of all fermion sectors are given by the 
form (1.1), then the Cabibbo-Kobayashi-Maskawa (CKM) matrix $V$ 
will be given by 
$V=V_TV_T^{\dagger}={\bf 1}$  (or $V=V_T^2$).   
We must assume an additional term which will break 
the cyclic permutation symmetry.  
Nevertheless, in this paper, we would like to emphasis that the mass 
matrix form (1.1) which is given on the cyclic permutation symmetric 
basis will shed a new light on the unified description of quarks and 
leptons.  The purpose of the present paper is not to give a theoretical 
model for the unified description, but to show how we can see many suggestive 
empirical relations for the fermion masses and mixings if we take a basis on 
which the mass matrices are cyclic permutation invariant.  

%%%%%%%%%%%%%%%%%%%%%%%%%%%%%%%%%%%%%%%%%%%%%%%%%%%%%%%
\section{Basic assumptions}

Our basic assumptions for the fermion mass matrices $M_f$ are as follows:
(i) The mass matrices $M_f$ are given by a bi-linear form
$$
M_f=m_{0}^{f}K_{f}K_{f}^{\dagger},
\eqno(2.1)
$$
and (ii) the matrices $K_f$ have a cyclic permutation invariant form
$$
K_f=a_{f}E-b_{f}S(\theta_{f}),
\eqno(2.2)
$$
where $E$ and 
$S$($\theta_f$) are defined in Eq.~(1.2).  
Here, in contrast to the definition of $M$ given in Eq.~(1.1), 
we have changed the sign of the coefficient of $S(\theta)$ such as 
the angle $\theta_{f}$ is in the range $0\leq\theta_f\leq\pi/2$ for 
$a_f>0$, $b_f>0$ and $m_1^f<m_2^f<m_3^f$.  
In the expression (2.2), the 
substantial parameters are only $b_f/a_f$ and $\theta_f$, because we discuss 
only the mass rations in the present paper.  

The form of $M_f$, (2.1), may be interpreted by a generalized seesaw 
scenario \cite{USMM}
$$
(\overline{f}_L\ \overline{F}_L)
\left(\begin{array}{cc}
0 & m_L \\
m_R & M_F
\end{array}\right) \left(
\begin{array}{c}
f_R \\
F_R
\end{array} \right) \ ,
\eqno(2.3)
$$
with $m_L{\propto}m_R{\propto}K_f$ and $M_F{\propto}{\bf 1}$, where $F$
 are hypothetical heavy fermions in addition to the conventional 
quarks and leptons $f$.  
However, in the present paper, we do not 
discuss the origin of the form of $M_f$ given in Eq.~(2.1) 
and confine ourselves to discuss phenomenological aspects 
of the model.  

Note that if we assume $b_f/a_f=1$, we obtain the relation \cite{Koide-me}
$$
m_1+m_2+m_3={\frac{2}{3}}({\sqrt{m_1}}+{\sqrt{m_2}}+{\sqrt{m_3}})^2,
\eqno(2.4)
$$
which is excellently satisfied by the observed charged lepton masses.  

In Table \ref{T-1}, we give values of $b_f/a_f$ and $\theta_f$ which are
 evaluated 
from the observed values of $m_i^f$ and the relations
$$
R\equiv3
{
\frac{m_1+m_2+m_3}
{({\sqrt{m_1}}+{\sqrt{m_2}}+{\sqrt{m_3}})^2}
}
=1+(b_f/a_f)^2,
\eqno(2.5)
$$
and 
$$
\tan\theta_f=
{\sqrt{3}}
{
\frac{{\sqrt{m_3}}-{\sqrt{m_2}}}
{{\sqrt{m_3}}+{\sqrt{m_2}}-2{\sqrt{m_1}}}       
}
\ .
\eqno(2.6)
$$
For comparison, we have given the values $b_f/a_f$ and $\theta_f$ for 
the following three cases: (a) the observed charged lepton masses and 
the running quark masses $m_i^q(\mu)$ at $\mu=m_Z$, (b) the fermion 
masses $m_i^f(\mu)$ at $\mu=\Lambda_X=2\times10^{16}$ GeV in a non-SUSY 
scenario, and (c) the fermion masses $m_i^f(\mu)$  at 
$\mu=\Lambda_X=2\times10^{16}$ GeV in a SUSY 
scenario (the quark mass values have been quoted from \cite{q-mass}).
As seen in Table \ref{T-1}, the values $b_f/a_f$ and $\theta_f$ 
are not so sensitive to the renomalization group effects 
(evolution of the Yukawa coupling constants), because those have 
been determined only from mass ratios.

We may read the values $\theta_f$ given in Table \ref{T-1} as
$$
\cos\theta_e
=
{\sqrt{\frac{11}{24}}},\ \ 
\cos\theta_d
=
{\sqrt{\frac{9}{24}}},\ \ 
\cos\theta_u
=
{\sqrt{\frac{7}{24}}},
\eqno(2.7)
$$
which give the angle values $\theta_e=47.39^{\circ}$, $\theta_d=52.24^{\circ}$
 and $\theta_u=57.31^{\circ}$, respectively, and give the relation 
$\cos^{2}\theta_{e}-\cos^{2}\theta_{d}=\cos^{2}\theta_{d}-\cos^{2}\theta_{u}
=1/12$.  
Of course, the relations (2.7) may be accidental coincidence, and they 
do not need to be taken seriously.

Note that not only the down-lepton masses, but also the down-quark masses give 
$b_f/a_f\simeq1$, so that the down-quark masses also satisfy the relation 
given in Eq.~(2.4).  
[However, the value of $m_1^f/m_2^f$ is sensitive to the deviation 
of $b_f/a_f$ from $b_f/a_f=1$.  If we take $b_d/a_d=1$, we must accept the 
prediction of $m_1^d$ with the deviation of $2\sigma$.]  
On the other hand, in contrast to the values of $b_e/a_e$ and 
$b_d/a_d$, the value of $b_u/a_u$ is considerably deviated from $b_u/a_u=1$.  
For a reference, in Table \ref{T-1}, we list values of $\tan^{-1}(b_f/a_f)$, 
where we put $a_f$ and $b_f$ as $a_f=\cos\phi_f$ and $b_f=\sin\phi_f$ since 
only the ratio $b_f/a_f$ is meaningful.  It is interesting that $\phi_f$
 show $\phi_u\simeq\theta_d$ and $\phi_d\simeq\theta_e$ and 
$\phi_e=45^{\circ}$.

\section{Neutrino mass matrix}
We assume that neutrino mass matrix $M_{\nu}$ is generated by the 
seesaw mechanism\cite{seesaw-nu}, i.e., 
$$
M_{\nu}\simeq-M_DM_R^{-1}M_D^T \ ,
\eqno(3.1)
$$
where $M_D$ and $M_R$ are Dirac and Majorana mass matrices, 
$\overline{\nu}_LM_D{\nu}_R$ and $\overline{\nu}_R^cM_R{\nu}_R$, 
respectively, where $\nu_R^c=(\nu_R)^c=C\nu_R^T$.  Since we consider 
that in a similar way to the charged lepton sector, the Dirac mass 
matrix $M_D$ is diagonalized by the tri-maximal mixing matrix 
$V_T$ as
$$
V_TM_DV_T^{\dagger}=D_D={\rm diag}(m_1^D,m_2^D,m_3^D) \ . 
\eqno(3.2)
$$
The neutrino mass matrix $M_\nu$ given in Eq.~(3.1) is written as
$$
M_{\nu}'=V_TM_{\nu}V_T^T \simeq -D_D(V_T^{\ast}M_RV_T^{\dagger})^{-1}
D_D \ .
\eqno(3.3)
$$
If the Majorana mass matrix $M_R$ is simply given by $M_R=m_R {\bf 1}$, 
then, by using the relation
$$
V_T V_T^T =
\left(\begin{array}{ccc}
1 & 0 & 0 \\
0 & 0 & 1 \\
0 & 1 & 0 
\end{array}\right) \ ,
\eqno(3.4)
$$
we obtain the form
$$
M_{\nu}'=-{\frac{1}{m_R}}D_D
\left(\begin{array}{ccc}
1 & 0 & 0 \\
0 & 0 & 1 \\
0 & 1 & 0 
\end{array}\right)^{-1}
D_D
=-{\frac{1}{m_R}}
\left(\begin{array}{ccc}
(m_1^D)^2 & 0 & 0 \\
0 & 0 & m_2^Dm_3^D \\
0 & m_2^Dm_3^D & 0
\end{array}\right), 
\eqno(3.5)
$$
When we take $(m_1^D)^2{\simeq}m_2^Dm_3^D$, we obtain $m^{\nu}_1
\simeq|m_2^{\nu}|=|m_3^{\nu}|=m_2^Dm_3^D/m_R$ and 
$\sin^{2}2\theta_{23}=1$.  Thus, we can obtain a natural explanation of 
the maximal mixing between 
$\nu_{\mu}$ and $\nu_{\tau}$ which is suggested by the atmospheric neutrino
 data \cite{nu-atm}.  However, the present oversimplified scenario 
$(M_R=m_R{\bf 1})$ cannot give $|{\Delta}m_{12}^2|\ll|{\Delta}m_{23}^2|$ 
$({\Delta}m_{ij}^2=(m_i^{\nu})^2-(m_j^{\nu})^2)$ and 
$\sin^{2}2\theta_{12}\neq0$ which are required for the explanation 
of the solar neutrino data \cite{nu-solar}. 
In order to give a realistic numerical result, we must take other terms 
in $M_R$ and $M_D$ into consideration. 

For the structure of $M_R$, an alternative scenario is also attractive : the 
Majorana mass tern $M'_R=V_T M_R V_T^T$ on the basis ${\nu}'_R=V_T\nu_R$ 
is again given by the cyclic permutation invariant form
$$
M'_R=m_{R}K_{R}.
\eqno(3.6)
$$
[However, $M'_R$ is invariant under the cyclic permutation not of $\nu_{Ri}$, 
but of ${\nu}'_R$.]  Since $M'_R$ must be symmetric, i.e., $(M'_R)^T=M'_R$, 
the angle parameter $\theta$ must be zero : 
$$
K_R=a_{R}E-b_{R}S(0).
\eqno(3.7)
$$
If we take a special case $b_R/a_R={\sqrt{2}}$, i.e.,
$$
K_R={\frac{a_R}{\sqrt{3}}}
\left(\begin{array}{ccc}
1 & -1 & -1 \\
-1 & 1 & -1 \\
-1 & -1 & 1 
\end{array}\right), 
\eqno(3.8)
$$
we obtain
$$
(M'_R)^{-1}=-
{\frac{1}{m_R}}
{\frac{{\sqrt{3}}}{2a_R}}
\left(\begin{array}{ccc}
0 & 1 & 1 \\
1 & 0 & 1 \\
1 & 1 & 0 
\end{array}\right),
\eqno(3.9)
$$
and 
$$
M'_{\nu}\simeq-D_D(M'_R)^{-1}D_D=
{\frac{1}{m_R}}
{\frac{{\sqrt{3}}}{2a_R}}
\left(\begin{array}{ccc}
0 & m_1^Dm_2^D & m_1^Dm_3^D \\
m_1^Dm_2^D & 0 & m_2^Dm_3^D \\
m_1^Dm_3^D & m_2^Dm_3^D & 0
\end{array}\right).
\eqno(3.10)
$$
The form of $M'_\nu$ in Eq.~(3.10) is already discussed 
in Ref.~\cite{nuKoide} 
for the case 
$b_{\nu}=-1/2$ in the ``democratic seesaw" model \cite{KFzp}
$$
M_f\simeq-m_{L}M_F^{-1}m_R,
\eqno(3.11)
$$
$$
m_L={\frac{1}{\kappa}}m_R=
{\frac{m_0}
{{\sqrt{m_{\tau}+m_{\mu}+m_e}}}}
{\rm diag}(
{\sqrt{m_e}},{\sqrt{m_{\mu}}},{\sqrt{m_{\tau}}}
),
\eqno(3.12)
$$
$$
M_F=m_F\{
\left(\begin{array}{ccc}
1 & 0 & 0 \\
0 & 1 & 0 \\
0 & 0 & 1 
\end{array}\right)
+b_f
\left(\begin{array}{ccc}
1 & 1 & 1 \\
1 & 1 & 1 \\
1 & 1 & 1 
\end{array}\right)\},
\eqno(3.13)
$$
and it is known \cite{nuKoide} that the case $b_{\nu}=-1/2$ 
gives the maximal mixing between $\nu_{\mu}$ and $\nu_{\tau}$.

Furthermore, if we consider a special case with $m_2^D=m_3^D$, 
which arises from $\theta_D=0$ in $M_D=K_{D}K_{D}^{\dagger}$, 
we obtain a simple neutrino mass matrix of the form
$$
M'_{\nu}=
\left(\begin{array}{ccc}
0 & x & x \\
x & 0 & y \\
x & y & 0 
\end{array}\right),
\eqno(3.14)
$$
where $x=({\sqrt{3}}/2a_{R}m_{R})m_{1}^{D}m_{3}^{D}$ and 
$x/y=m_1^D/m_3^D$.  This matrix form given in Eq.~(3.14) has 
recently been proposed by Ghosal \cite{Ghosal} on the basis 
of discrete $Z_3{\times}Z_4$ symmetries.  
The mass matrix $M'_\nu$ given in Eq.~(3.14) gives the eigenvalues 
$(m_1^{\nu}, -m_2^{\nu}, -m_3^{\nu})$,
$$
m_{1}^{\nu}=
{\frac{1}{2}}(
{\sqrt{y^2+8x^2}}+y),\ \ 
-m_{2}^{\nu}=-
{\frac{1}{2}}(
{\sqrt{y^2+8x^2}}
-y),\ \ 
-m_{3}^{\nu}=-y,
\eqno(3.15)
$$
(we have defined $m_i^{\nu}$ as $m_i^{\nu}>0$ and 
$m_1^{\nu}>m_2^{\nu}>m_3^{\nu}$), and the mixing matrix 
$$
U=
\left(\begin{array}{ccc}
c_{12} & s_{12} & 0 \\
-{\frac{1}{{\sqrt{2}}}}s_{12} & {\frac{1}{{\sqrt{2}}}}c_{12} 
& -{\frac{1}{{\sqrt{2}}}} \\
-{\frac{1}{{\sqrt{2}}}}s_{12} & {\frac{1}{{\sqrt{2}}}}c_{12} 
& {\frac{1}{{\sqrt{2}}}}
\end{array}\right),
\eqno(3.16)
$$
where $c_{12}=\cos\theta_{12}={\sqrt{m_1^{\nu}/(m_1^{\nu}+m_2^{\nu})}}$ 
and $s_{12}=\sin\theta_{12}={\sqrt{m_2^{\nu}/(m_1^{\nu}+m_2^{\nu})}}$.  
Therefore, for $x{\gg}y$ $(|m_1^D|{\gg}|m_3^D|)$, 
we obtain the relations 
${\Delta}m_{12}^2\simeq2{\sqrt{2}}xy$ and ${\Delta}m_{23}^2\simeq2x^2$ 
together with $c_{12}{\simeq}s_{12}{\simeq}1/{\sqrt{2}}$, which leads to 
a bi-maximal mixing.  Thus, we can give a reasonable explanation both 
for the atmospheric and solar neutrino data regarding 
${\Delta}m_{12}^2{\simeq}{\Delta}m_{solar}^2{\sim}10^{-10}$ eV$^2$ and 
${\Delta}m_{23}^2{\simeq}{\Delta}m_{atm}^2{\sim}10^{-3}$ eV$^2$.  

%%%%%%%%%%%%%%%%%%%%%%%%%%%%%%%%%%%%%%%%%
\section{CKM mixing matrix}

If we apply the matrix form given in Eq.~(2.2) to the up- and down-quark sectors, 
we obtain the wrong result $V={\bf 1}$.  
Therefore, let us modify the matrix $K_f$ by adding a small term
$c_f P_\omega$:
$$
K_f=a_{f}E-b_{f}S(\theta_f)+c_{f}P_{\omega}\ .
\eqno(4.1)
$$
$$
P_{\omega}=
{\frac{1}{\sqrt{3}}}
{\rm diag}(\omega,\omega^2,1)\ .
\eqno(4.2)
$$
Then, the form (4.1) is not invariant under the cyclic permutation
$(f_1 \rightarrow f_2, f_2 \rightarrow f_3, f_3 \rightarrow f_1)$.
The term $(\omega \overline{f}_{1L} f_{1R} +
\omega^2 \overline{f}_{2L} f_{2R} + \overline{f}_{3L} f_{3R})$ is
transformed into $\omega^2 (\omega \overline{f}_{1L} f_{1R} +
\omega^2 \overline{f}_{2L} f_{2R} + \overline{f}_{3L} f_{3R})$
under the cyclic permutation.
However, if the $E$ and $S$ terms are absent in the matrix $K_f$,
the form of $P_\omega$ is invariant under the cyclic
permutation, because the common phase factor is unphysical.
We regard the small $P_\omega$ term as a ``form" invariant
term under the cyclic permutation in the extended meaning.

The modified matrix $K_f$ given in Eq.~(4.1), is transformed 
by the tri-maximal 
mixing matrix $V_T$ as 
$$
K_f'=V_TK_fV_T^{\dagger}
={\frac{1}{\sqrt{3}}}
\left(\begin{array}{ccc}
\lambda_1^f & 0 & 0 \\
0 & \lambda_2^f & 0 \\
0 & 0 & \lambda_3^f
\end{array}\right)
+
{\frac{c_f}{\sqrt{3}}}
\left(\begin{array}{ccc}
0 & 1 & 0 \\
0 & 0 & 1 \\
1 & 0 & 0 
\end{array}\right),
\eqno(4.3)
$$
where $\lambda_i^f$ are the eigenvalues of the matrix 
${\sqrt{3}}[a_{f}E -b_{f}S(\theta_f)]$ and 
they are explicitly given in Eq.~(1.5) with 
$m_i\rightarrow\lambda_i/{\sqrt{3}}$.  

The mixing matrix $U_L$ is obtained by the diagonalization of the 
Hermitian matrix 
$$
M'_f=m_0 K'_f (K'_f)^{\dagger}
={\frac{1}{3}}
m_0\left[
\left(\begin{array}{ccc}
\lambda_1^2 & \lambda_{2}c_f & \lambda_{1}c_f^{\ast} \\
\lambda_{2}c_f^{\ast} & \lambda_2^2 & \lambda_{3}c_f \\
\lambda_{1}c_f & \lambda_{3}c_f^{\ast} & \lambda_3^2 
\end{array}\right)
+ |c_f|^{2}{\bf 1}
\right].
\eqno(4.4)
$$
For a small $c_f$ and $\lambda_3^2\gg\lambda_2^2\gg\lambda_1^2$, the 
eigenvalues of $M'_f$ are given by
$$
m_1{\simeq}{\frac{1}{3}}m_0\lambda_1^2\ , \ \ 
m_2{\simeq}{\frac{1}{3}}m_0(\lambda_2^2+|c_f|^2)\ , \ \ 
m_3{\simeq}{\frac{1}{3}}m_0(\lambda_3^2+2|c_f|^2)\ , 
\eqno(4.5)
$$
and the mixing angles are given by
$$
\tan2\theta_{12}\simeq
{\frac{2\lambda_2|c_f|}{\lambda_2^2-\lambda_1^2}}
\simeq
2{\frac{|c_f|}{\lambda_2}},
\eqno(4.6)
$$
$$
\tan2\theta_{23}\simeq
{\frac{2\lambda_3|c_f|}{\lambda_3^2-\lambda_1^2}}
\simeq
2{\frac{|c_f|}{\lambda_3}},
\eqno(4.7)
$$
and $\theta_{13} \simeq 0$. (If $c_f$ is complex, i.e., arg$c_f\neq 0$, the 
elements $U_{12}=\sin\theta_{12}$ and $U_{23}=\sin\theta_{23}$ are replaced 
by $U_{12}=\sin\theta_{12} \exp (i \arg c_f)$ and $U_{23}=\sin\theta_{23}
 \exp (i \arg c_f)$, respectively.)  
If we consider that the value of $|c_f|^2$ 
in the up-quark sector is sufficiently small so that 
$\theta_{ij}^u{\simeq}0$, we obtain
$$
\left|
{\frac{V_{cb}}{V_{us}}}
\right|
\simeq
{\frac{|U_{23}^d|}{|U_{12}^d|}}
\simeq
{\frac{\lambda_2^d}{\lambda_3^d}}
\simeq
\sqrt{\frac{m_s}{m_b}}
=0.176
,
\eqno(4.8)
$$
which is in good agreement with the experimental value \cite{pdg} 
$|V_{cb}/V_{us}|=(0.0373\pm0.0018)/(0.2205\pm0.0018)=0.169\pm0.008$.  
Note that in order to obtain the relation (4.8), the assumption (i) 
given in Eq.~(2.1) is essential.

When we denote $M'_f$ as
$$
M'_f = M_f^0 + \frac{1}{3}|c_f|^2 {\bf 1} \ ,
\eqno(4.9)
$$
as a trial, let us suppose that the lowest one of the three eigenvalues 
of the mass matrix $M_f^0$ except for the common constant term 
$|c_f|^2${\bf 1} is 
assigned to zero, i.e., 
$$
{\rm det} M_f^0 = 0 \ . 
\eqno(4.10)
$$
In this scenario, we obtain
$$
|c_f|^2\simeq\lambda_1^2,  
\eqno(4.11)
$$
together with $m_1^f = |c_f|^2/3$.
Neglecting the mixings in the up-quark sector, we obtain
$$
|V_{us}|\simeq |c_d|/\lambda_2^d\simeq
\sqrt{m_d/m_s}
=0.224,
\eqno(4.12)
$$
$$
|V_{cb}|\simeq |c_d| \lambda_3^d\simeq
\sqrt{m_d/m_b}
=0.0395, 
\eqno(4.13)
$$
which are in good agreement with the experiments.  However, at present, 
there is no theoretical understanding behind the assumption given in
Eq.~(4.10).

The existence of the $P_{\omega}$ term with $|c_f|^2{\simeq}(\lambda_1^f)^2$ 
slightly changes the predicted values of $m_i^f$ from the case of $c_f=0$.  
Especially the values of $m_2^f$ is sizably changed as 
$m_2^f=(\lambda_2^f)^2/3{\rightarrow}m_2^f{\simeq}[(\lambda_2^f)^2+|c_f|^2]/3$.  However, for the charged leptons, the existence dose not so 
badly spoil the relation given in Eq.~(2.4), because 
$|c_{e}|^{2}/(\lambda_{2}^{e})^{2}{\simeq}(\lambda_{1}^{e}/\lambda_{2}^{e})^{2}{\simeq}(m_{1}^{e}/m_{2}^{e}){\sim}10^{-3}$.  In order to 
obtain the best fit values of $m_i^f$, the values of $b_f/a_f$ and $\theta_f$ 
will slightly be changed from the values given in Table 1.  

%%%%%%%%%%%%%%%%%%%%%%%%%%%%%%%%%%%%%%%%%%%%%
\section{Concluding remarks}

In conclusion, we have pointed out that we can see many
interesting relations for the quark and lepton masses
and their mixings if we take a basis on which the form of the mass
matrices is invariant under the cyclic permutation 
$(f_1{\rightarrow}f_2, f_2{\rightarrow}f_3, f_3{\rightarrow}f_1)$.
We have assumed that the mass matrices $M_f$ are given by the
bi-linear form [given in Eq.~(2.1)], where the matrix $K_f$ is given 
by Eq.~(4.1).
If we put $b_e/a_e=1$ together with $c_e=0$, we obtain the
relation given in Eq.~(2.4) which is in excellent agreement 
with experiments.
We think that the form of the down-lepton mass matrix
with $b_f/a_f=1$ is the most fundamental one.
Although there is at present no explicit theoretical ground 
for the ansatz given in Eq.~(4.10), however,
if we accept the ansatz, we can obtain the
well-satisfied relations for $|V_{us}|\simeq{\sqrt{m_d/m_s}}$ 
and $|V_{cb}|\simeq{\sqrt{m_d/m_b}}$.
It is also worth while to note that the model naturally leads to 
the maximal mixing between $\nu_\mu$ and $\nu_\tau$.
(This result is independent of the assumption given in Eq.~(2.1).)
If we take the special values of the parameters, we can
obtain a beautiful neutrino mass matrix (Ghosal's neutrino
mass matrix) given in Eq.~(3.14) which leads to  bi-maximal mixing.

%%%%%%%%%%%%%%%%%%%%%%%%%%%%%%%%%%%%%%%%%%%
\vspace*{.3in}

\centerline{\Large\bf Acknowledgments}

This work started from enjoyable conversation with
Prof.~E.~Takasugi. 
The author would like to thank Professor E.~Takasugi for
his exciting discussions and informing valuable references
to the author. 
He also thank Dr.~A.~Ghosal for  helpful discussions and 
useful comments.

%\newpage
%%%%%%%%%%%%%%%%

%%%%%%%%%%%%%%%%%

%%%%%%%%%%%%%%%%%%%%%%%%%%%%%%%%%%%%%%%%%%%%%%%%%%%%
%\newpage
\vspace{.3in}

\begin{table}
\caption{Numerical values of the parameters $b_f/a_f$ [and also
$\phi_f=\tan^{-1}(b_f/a_f)$] and $\theta_f$,
which are defined by (2.5) and (2.6) in the text.  
Input values $m_i^f(\mu)$ are 
quoted from Ref.~[4], where $\Lambda_X=2\times10^{16}$ GeV.
The values of $b_f/a_f$ and $\theta_f$ have been evaluated only from
the center values of $m_i^f(\mu)$.
\label{T-1}
}

\vglue.1in

\begin{tabular}{|l|ccc|ccc|}\hline
  & \multicolumn{3}{c|}{Inputs} & \multicolumn{3}{c|}{Outputs} \\
  & $m_1^f$(MeV) & $m_2^f$(MeV) & $m_3^f$(GeV) & $b_f/a_f$ & $\phi_f$ & 
$\theta_f$ \\ \hline
$(m_i^e)_{obs}$ & $0.510999$ & $105.6584$ & 
$1.77705^{+0.00029}_{-0.00029}$ & 
$1.0000$ & $45.000^{\circ}$ & $47.2680^{\circ}$  \\
$m^e_i{(\Lambda_X)}_{non-SUSY}$ & $0.493486$ & $104.152$ & 
$1.7706{\pm}0.0003$ & $1.0019$ & $45.054^{\circ}$ & $47.336^{\circ}$ \\
$m_i^e(\Lambda_X)_{SUSY}$ & $0.3250203$ & $68.598$ & 
$1.1714{\pm}0.0002$ & $1.0026$ & $45.073^{\circ}$ & $47.367^{\circ}$ \\ \hline
$m^d_i(m_Z)$ & $4.69^{+0.60}_{-0.66}$
 & $93.4^{+11.8}_{-13.0}$
 & 
$3.00\pm0.11$ & 
$1.047$ & $46.30^{\circ}$ & $52.43^{\circ}$ \\
$m^d_i(\Lambda_X)_{non-SUSY}$ & $1.49^{+0.25}_{-0.28}$
 & 
$38.7^{+4.9}_{-5.4}$
 & 
$1.07\pm0.04$ & 
$1.024$ & $45.69^{\circ}$ & $51.77^{\circ}$ \\
$m^d_i(\Lambda_X)_{SUSY}$ & $1.33^{+0.17}_{-0.19}$
 & 
$26.5^{+3.3}_{-3.7}$
 & 
$1.00\pm0.04$ & 
$1.070$ & $46.93^{\circ}$ & $53.07^{\circ}$ \\ \hline
$m^u_i(m_Z)$ & $2.33^{+0.42}_{-0.45}$
 & 
$677^{+56}_{-61}$
 &
$181\pm13$ &
$1.287$ & $52.15^{\circ}$ & $57.05^{\circ}$ \\
$m^u_i(\Lambda_X)_{non-SUSY}$ & $0.94^{+0.17}_{-0.18}$
 & 
$272^{+22}_{-24}$
 & 
$84^{+18}_{-13}$
 & 
$1.295$ & $52.33^{\circ}$ & $57.26^{\circ}$ \\
$m^u_i(\Lambda_X)_{SUSY}$ & $1.04^{+0.19}_{-0.20}$
 & 
$302^{+25}_{-27}$
 & 
$129^{+196}_{-40}$
 & 
$1.312$ & $52.68^{\circ}$ & $57.68^{\circ}$ \\ \hline
\end{tabular}
\end{table}

\end{document}